\documentclass[twocolumn,preprintnumbers,secnumarabic,amsmath,amssymb,nofootinbib,floatfix]{revtex4}
\usepackage{varioref,exscale,latexsym,amsmath,amssymb}
\usepackage{graphicx}

\usepackage{graphicx}
\usepackage{dcolumn}
\usepackage{bm}
\usepackage{slashed}

\def\beq{\begin{equation}}

\def\eeq{\end{equation}}

\def\beqa{\begin{eqnarray}}

\def\eeqa{\end{eqnarray}}

\begin{document}
\title{{\bf Spatial variation of fundamental couplings and Lunar Laser Ranging \\}}
\medskip
\author{Thibault Damour$^{a}$}
\author{ John F. Donoghue$^{a,b}$}
\affiliation{${}^a$Institut des Hautes \'{E}tudes Scientifiques \\
Bures sur Yvette, F-91440, France\\
and \\
${}^b$Department of Physics\\
University of Massachusetts\\
Amherst, MA  01003, USA
}
\date{15 April 2011}

\begin{abstract}
If the fundamental constants of nature have a cosmic spatial variation, there will in general be extra forces with a preferred direction in space which violate the equivalence principle. We show that the millimeter-precision Apache Point Observatory Lunar Laser-ranging Operation
 provides a very sensitive probe of such variation that has the capability of detecting a
 cosmic gradient of  the ratio between the quark masses and the strong interaction scale at the level
 $\mathbf{\nabla} \ln (m_{\rm quark}/\Lambda_{QCD}) \sim  2.6 \times 10^{-6} ~{\rm Glyr}^{-1}$, which is comparable to the cosmic gradients suggested by the recently reported measurements of Webb et al. We also point out the capability of presently planned improved equivalence principle
 tests, at the $\Delta g/g \lesssim 10^{-17}$ level, to probe similar cosmic gradients.
 \end{abstract}
\maketitle

\section{Introduction}

Within many extensions of the Standard Model, the parameters of our fundamental theory need not be universally constant but may vary in space and time.
The search for such variations provides important constraints on such theories. Recently, Webb et al. \cite{Webb:2010hc} have reported evidence for a
non-zero spatial variation of the fine structure constant $\alpha$. Parameterizing the variation
of $\alpha$ by a dipole gradient
\begin{equation} \label{dipolealpha}
\frac{\alpha(\mathbf{x})}{\bar{\alpha}} = 1 + B_\alpha \,  \mathbf{\hat{z}}_\alpha \cdot \mathbf{x} 
\end{equation}
they find evidence, at the $4.2 \,  \sigma$ level, for a slope parameter
\begin{equation} \label{Balpha}
B_\alpha = (1.10 \pm 0.25) \times 10^{-6} ~ {\rm Glyr}^{-1}
\end{equation}
relative to the unit direction $\mathbf{\hat{z}}_\alpha$ of right ascension $\alpha=17.4 \pm 0.6$ hours  and declination $\delta=-58 \pm 6 $ degrees.
In addition, Berengut et al. \cite {Berengut:2010yu} found weak indications for the existence of a gradient of the electron to proton mass ratio $\mu \equiv m_e/m_p$
in the same  direction $\mathbf{\hat{z}}_\mu = \mathbf{\hat{z}}_\alpha$, with slope

\begin{equation} \label{Bmu}
B_\mu = (2.6 \pm 1.3) \times 10^{-6} ~ {\rm Glyr}^{-1} \, .
\end{equation}
Other spatial gradients are much more weakly tested. For example, Donoghue and Donoghue \cite{Donoghue:2004gu} have used the
spatial constancy of the first acoustic peak in the Cosmic Microwave Background to bound a possible
variation in the cosmological constant (or generalized dark energy) at the level of an analogous slope parameter
\begin{equation}
B_\Lambda < 0.91 \times 10^{-2} ~ {\rm Glyr}^{-1}
\end{equation}
at the 95\% confidence level.

Because the masses of all the elements depend on the parameters of the Standard Model, a gradient in one
of these parameters will lead to a force (as noted long ago by Dicke \cite{Dicke}).
Using the fine structure constant as an example,  the dependence on $\alpha$
of the total mass-energy of system $A$,
\begin{equation}
E_A(\alpha) = c^2 M_A(\alpha )  ~~,
\end{equation}
implies that a spatial gradient  $\mathbf{\nabla} \alpha$ of $\alpha$ ,
 will lead to a force
\begin{equation}
\mathbf{F}= -\mathbf{\nabla} E_A (\alpha) = - c^2  \frac{\partial M_A}{\partial \alpha} \mathbf{\nabla} \alpha   ~~,
\end{equation}
If we introduce the following dimensionless effective ``charge'' associated to the $\alpha$ dependence,
\begin{equation}
Q_{\alpha}(A) = \frac{\alpha}{M_A}\frac{\partial M_A}{\partial \alpha}
\end{equation}
and parameterize the gradient of $\alpha$ by a slope and a unit direction as in Eq. (\ref{dipolealpha}),
 $\mathbf{\nabla} \alpha/ \alpha =  B_\alpha \,  \mathbf{\hat{z}}_\alpha$, the above force reads
\begin{equation}
\mathbf{F}_A  = -Q_{\alpha}(A)~ M_A ~B_\alpha c^2~\mathbf{\hat{z}}_\alpha \, .
\end{equation}

If we now consider the dependence of the total mass-energy $ M_A c^2$ (in units of the Planck mass) of system $A$
on the various dimensionless ratios (or  coupling constants) $r_i = \alpha, \mu, m_{\rm quark}/m_p, \ldots$
entering physics at energy scales $\lesssim m_p c^2$,  and if we assume the existence of (fractional) spatial
gradients $\mathbf{\nabla} \ln r_i =  B_{r_i} \,  \mathbf{\hat{z}}_{r_i}$ of the various dimensionless
ratios, we see that body $A$ will be submitted to an external {\it acceleration},
$\mathbf{g}_A $,  of the form
\begin{equation} \label{gA}
\mathbf{g}_A = \frac{\mathbf{F}_A}{M_A}  = - \sum_i Q_{r_i}(A) ~B_{r_i} c^2~\mathbf{\hat{z}}_{r_i}
\end{equation}
where $Q_{r_i} = Q_\alpha, Q_\mu, \ldots$ are the various dimensionless effective ``charges''
associated to the dependence of the mass on the various ratios (or coupling constants), namely
\begin{equation} \label{defQ}
Q_{r_i}(A) \equiv  \frac{r_i}{M_A}\frac{\partial M_A}{\partial r_i} = \frac{\partial \ln (M_A/M_P)}{\partial \ln r_i}
\end{equation}
In the second form of the definition of  $Q_{r_i}$ we have recalled that $M_A$ is to
be expressed in units of the Planck mass $M_P$. [This corresponds to working in the
``Einstein conformal frame", where Newton's constant is held fixed.]

 If the various effective charges $Q_{r_i}(A)$ were independent of the considered
 body $A$, the result would be an unobservable (gravity-like) uniform free fall with a universal
 acceleration  $\mathbf{g}_0=\mathbf{g}_A=\mathbf{g}_B$ in a direction given by
 an average of the various gradients $\mathbf{\nabla} \ln r_i =  B_{r_i} \,  \mathbf{\hat{z}}_{r_i}$.
 However, composition dependence of (at least one of) the various charges, e.g.  $Q_{\alpha}(A) -Q_{\alpha}(B) \ne 0$,  or  $Q_{\mu}(A) -Q_{\mu}(B) \ne 0$, will lead to differential accelerations
$ \mathbf{g}_A -\mathbf{g}_B \ne 0$ and locally observable effects.
Recently, we have studied \cite{Damour:2010rm, Damour:2010rp}
the composition dependence of the
effective charges $Q_{r_i}(A)$ corresponding to a complete set of dimensionless
ratios entering low-energy physics, namely $r_0= \Lambda_{QCD}/M_P$,
and
\begin{equation} \label{defri}
 r_i = \alpha, m_u/\Lambda_{QCD}, m_d/\Lambda_{QCD}, m_e/\Lambda_{QCD} \, ,
 \end{equation}
where $m_u, m_d$ are the masses of the light quarks.\footnote{We have argued in \cite{Damour:2010rm, Damour:2010rp} that the composition dependence linked to the strange quark mass
$m_s$ was subdominant.} Here, we separated the ratio $r_0= \Lambda_{QCD}/M_P$,
the dependence on which leads to composition-independent effects. We shall recall below
our explicit results for the various charges $Q_{r_i}(A)$ ($i \ne 0$).

Each slope parameter $ B_{r_i}$ defines a corresponding
acceleration $  B_{r_i} c^2 $,  which enters the total acceleration (\ref{gA}), multiplied
by the corresponding dimensionless effective charge  $Q_{r_i}(A)$. For instance,
the $\alpha$ gradient  (\ref{Balpha}) reported by Webb et al. \cite{Webb:2010hc} corresponds
(using $c/1$yr$=950 \, $cm/s$^2$) to the acceleration level
\begin{equation} \label{Balphac2}
 B_\alpha c^2 =(1.05 \pm 0.24) \times 10^{-12} \, {\rm cm/s}^2
\end{equation}
while the acceleration level corresponding to the $\mu$ gradient
suggested by  Berengut et al. \cite {Berengut:2010yu} is
\begin{equation} \label{Bmuc2}
 B_\mu c^2 =(2.5 \pm 1.2)  \times 10^{-12} \, {\rm cm/s}^2
\end{equation}

The aim of this paper is to point out that  the EP-violating effects of
spatial gradients of $\alpha$ and of the mean quark-mass ratio\footnote{Note that, from a theoretical point of view, the
ratio  $r_{\hat{m}} = \hat{m}/\Lambda_{QCD}$ is  akin to $\mu=m_e/m_p$ in the
sense that it is the ratio between a lepton mass and an hadronic one.}
$r_{\hat{m}} = \hat{m}/\Lambda_{QCD}$ (with $\hat{m} \equiv (m_u+m_d)/2$) at the levels  Eqs. (\ref{Balphac2}), (\ref{Bmuc2}) generate  signals  in the ranging to the Moon, which have a specific time structure, and an amplitude which seems large enough to be detectable by
the recently started millimeter-precision Apache Point Observatory Lunar Laser-ranging Operation (APOLLO)
\cite{Murphy:2007ed,apollo09}. [See \cite{lunarlaser} for the
results obtained from the pre-APOLLO  Lunar Laser Ranging (LLR) experiments.] We also describe the weaker bounds on spatial gradients obtained
 by present laboratory-based experiments \cite{Schlamminger:2007ht}, and indicate that
 planned EP experiments at the $\Delta g/g \lesssim 10^{-17}$ level will probe
 cosmic gradients at the levels of Eqs. (\ref{Balphac2}), (\ref{Bmuc2}).

Let us here emphasize the difference in outlook between our previous work, and the
present study. In Refs.~ \cite{Damour:2010rm, Damour:2010rp}, we were considering the
case where the spatial or temporal variation of a dimensionless parameter indicates the existence of a field, say $\varphi$, which carries the spacetime dependence, and we were considering the violations
of the ``weak version'' of the Equivalence Principle (EP), i.e. composition-dependent accelerations of body $A$,
mediated by the coupling of $\varphi$  to {\it local} matter distributions.  As a consequence, the locally observable
EP-violating effects depended on the product of {it two} $\varphi$ coupling strengths, say
$\alpha_A \, \alpha_E$, where
$$\alpha_A = \partial \ln (M_A(\varphi)/M_P)/ \partial \ln \varphi$$
measures the coupling of $\varphi$ to body $A$, and $\alpha_E = \partial \ln (M_E(\varphi)/M_P)/ \partial \ln \varphi$ its coupling to an external ``source''  body $E$ (which could be the Earth, the Sun, or some laboratory source).  However, we had normalized the definition of the fundamental
couplings $d_{r_i}$ of the ``dilaton''  field $\varphi$,
as they enter the low-energy Lagrangian, so that we could write each $\alpha_A $ in the
specific form
\begin{equation} \label{alphaA}
\alpha_A = d_{r_0} + \sum_{i\ne0}  d_{r_i} Q_{r_i}(A)
\end{equation}
exhibiting a simple factorization between the fundamental dilaton couplings  $d_{r_0}, d_{r_i}$,
and the phenomenological effective charges  defined in Eq. (\ref{defQ}) above. Note that there is
no composition-dependent charge associated to the coupling to $r_0= \Lambda_{QCD}/M_P$,
or, said differently, the charge $Q_{r_0}(A)$ associated to $r_0= \Lambda_{QCD}/M_P$
is simply $Q_{r_0}(A) \equiv 1$
because, as the mass  $M_A$ can be written as the product of the hadronic mass scale $\Lambda_{QCD}$
by a dimensionless function $ f(r_i)$ of the dimensionless ratios,  Eq.~(\ref{defri}), the mass ratio
$M_A/M_P$ can be identically written as $M_A/M_P= r_0 f(r_i)$.
 [Note also that the
$ d_{r_i}$'s entering  Eq. (\ref{alphaA}) above correspond to the differences $d_{r_i} - d_g$ in
 \cite{Damour:2010rm, Damour:2010rp}  (with $d_g \equiv d_{r_0}$), because we defined above the
 ratios $r_i$ by Eq. (\ref{defri}) which involved a logarithmic derivative of $M_A/\Lambda_{QCD}$, while we were
 working there with logarithmic derivatives of $M_A/M_P$.]

When contemplating, as we do here, possible variations over cosmological distances,
the field $\varphi$ must be essentially massless. However, in the present work we
shall not need to consider any specific model neither for the mass (or self-potential
$V(\varphi)$) of $\varphi$, nor for its matter couplings $d_{r_0}, d_{r_i}$. Indeed, the
crucial point is that the observable acceleration (\ref{gA}) only depends on the
effective charges  (\ref{defQ}), and on the various spatial gradient parameters
$\mathbf{\nabla} \ln r_i =  B_{r_i} \,  \mathbf{\hat{z}}_{r_i}$. This makes the present investigation
quite model independent, as well as independent from the usual interpretation of local
EP tests (which involve the bilinear products  $(\alpha_A - \alpha_B) \, \alpha_E$).


\section{The gravitational Stark effect and spatial varying couplings}

The accurate monitoring of the lunar motion (most notably by LLR experiments \cite{lunarlaser}) has led to impressive
tests of relativistic gravity, and notably of various aspects of the EP  \cite{Nordtvedt:1968zz}.
Here we are interested in EP-violating effects in LLR that are linked to a {\it fixed} preferred direction in space.  Such effects have been studied by Damour and Schaefer  \cite{Damour:1991rq}
in the context of binary pulsars. The analysis of such preferred-direction forces in the context
of LLR has been done at leading order (LO) by Nordtvedt \cite{Nordtvedt94} (see also
 \cite{Muller:1996up}) , and to very high perturbative order
by Damour and Vokrouhlick\'{y} \cite{Damour:1995zr}  (using  the Hill-Brown lunar theory
\cite{BrouwerClemence}).
For references to analytic studies of relativistic effects in lunar
motion, as well as a self-contained introduction to Hill-Brown theory, see, e.g.,  \cite{Damour:1995gi}.
Lunar dynamics is a notoriously difficult problem because of the rather
strong perturbation coming from the Sun's tidal forces, which leads to badly convergent perturbation
series in powers of the  parameter $m=n'/(n-n') \simeq 1/12.3687$. [Here, $n' = 2 \pi/(1~{\rm yr})$ denotes the mean sidereal angular velocity
of the Earth around the Sun, and $n=  2\pi/(27.32~ {\rm days})$  \cite{book}  the mean sidereal angular velocity of the Moon around the Earth.]
For some effects, a LO perturbation treatment in $m$ can be significantly inaccurate both because of the occurrence of small denominators, and of the slow
convergence of the $m$-perturbation series. This is, for instance, the case for the Laplace-Nordtvedt effect of polarization
of the Moon's orbit by an EP-violation linked to the Sun's gravity where higher-order terms in $m$ increase
the leading-order result by more than $62 \%$ \cite{Nordtvedt95,Damour:1995gi}.
In the case of interest here of what has been called the ``gravitational Stark effect'' \cite{Damour:1991rq}, i.e. the perturbing  influence of a differential force (with a {\it fixed direction}) acting on a gravitationally bound
two-body system, the situation is similar, though with significant differences.

Let us first recall that the classical (electric or gravitational) Stark effect is an example of singular perturbation where a small perturbing force can have a large effect.
If we were approximating the dynamics of the relative Earth-Moon coordinate $\mathbf{r}= \mathbf{x_M}-\mathbf{x_E}$ in the presence of an external acceleration $\mathbf{\Delta \mathbf{g}}=
 \mathbf{g}_M -  \mathbf{g}_E$ by means of the Lagrangian
 [after factorization of the Earth-Moon reduced mass $\mu \equiv m_E \, m_M / (m_E + m_M)$]
\begin{equation}
\label{2.1}
L = \frac{1}{2} \left( \frac{d {\bm r}}{dt} \right)^2 + \frac{G(m_E + m_M)}{r} + \bm{\Delta g} \cdot {\bm r} \, ,
\end{equation}
we could find the exact solution of the perturbed dynamics by separating the Hamilton-Jacobi equation corresponding to the Lagrangian (\ref{2.1}) in parabolic coordinates ($\xi = r+z$, $\eta = r-z$, $\phi$), with a $z$ axis oriented along $\bm{\Delta g}$. One then finds that the exact solution corresponding to elliptic orbits undergoes a complicated secular evolution during which the osculating elements of the elliptic motion wander very far away from any given initial state. For instance, even if the perturbing acceleration $\bm{\Delta g}$ is very small, the osculating eccentricity will not undergo small oscillations around its initial value $e_0$ but will, on time scales $na / \Delta g$, take values quite different from $e_0$. This instability of elliptic motion under a constant force can also be seen by using the {\em averaged} evolution equations of the osculating orbital elements. More precisely, if we consider the evolution of the semi-major axis $a$, of the Lagrange-Laplace-Runge-Lenz
eccentricity vector $\mathbf{e} = e\mathbf{a}$ (directed towards the periastron), where $(\mathbf{a},~\mathbf{b}, ~\mathbf{c}$) are orthornormal
unit vectors with $\mathbf{a}$ pointing towards the periastron and $\mathbf{c}$ along the orbital angular momentum $\bm{\ell} =  (1-e^2)^{1/2}\mathbf{c} =\mathbf{r}\times \mathbf{v}$, one finds
averaged evolution equations of the form~\cite{Damour:1991rq}
\begin{eqnarray} \label{secular}
\langle\frac{da}{dt}\rangle &=& 0 ~~, \nonumber \\
\langle\frac{d\mathbf{e}}{dt}\rangle &=& \mathbf{f}\times \bm{\ell} ~~, \nonumber\\
\langle\frac{d\bm{\ell}}{dt}\rangle &=& \mathbf{f}\times \mathbf{e}
\end{eqnarray}
where
\begin{equation}
\mathbf{f} = \frac{3}{2}\frac{\Delta\mathbf{g}}{na}~~.
\end{equation}
with  $n$ denoting the sidereal angular frequency of the Moon.

We see that while $a$ stays secularly constant, the vectors ${\bm e}$ and ${\bm \ell}$ rotate one into another. More precisely (with ${\bm f} = f  \, \mathbf{\hat{z}}$), one easily sees that,
while $e_z$ and $\ell_z$ stay constant, the two complex combinations $\varepsilon_x \equiv e_x + i \, \ell_y$ and $\varepsilon_y \equiv e_y + i \, \ell_x$ rotate as
\begin{equation}
\label{2.2}
\varepsilon_x (t) = e^{+ift} \, \varepsilon_x (0) \, ; \quad \varepsilon_y (t) = e^{-ift} \, \varepsilon_y (0) \, ,
\end{equation}
leaving constant $\vert \varepsilon_x \vert^2 = e_x^2 + \ell_y^2$ and $\vert \varepsilon_y \vert^2 = e_y^2 + \ell_x^2$ (consistently with ${\bm e}^2 + {\bm \ell}^2 = 1 = {\rm const.}$).

\smallskip

This Stark instability is rooted in the  well known degeneracy of the Coulomb problem, {\em i.e.} the fact that the radial $\omega_r$ and angular frequencies $\omega_{\phi}$ happen to be exactly equal, $\omega_r = \omega_{\phi} = n$, for a $1/r$ interaction potential. As a consequence, any lifting of the Coulomb degeneracy by an additional interaction potential (causing $\omega_r$ to differ from $\omega_{\phi}$) will tame the Stark instability. Ref.~\cite{Damour:1991rq} considered the case where this lifting was due to the general relativistic modifications of the $1/r$ Newtonian potential. In that case the only modification of the secular evolution equations Eqs.~(\ref{secular}) is the appearance of an additional
contribution $+ \,  \omega_p \, \mathbf{c}\times\mathbf{e}$ on the r.h.s. of the evolution
equation of  $\mathbf{e}$, where $\omega_p= \omega_{\phi} - \omega_r$ is the precession frequency of the
binary system, due to relativistic effects.

As a first orientation towards understanding the Stark effect in the lunar motion,  let us start
by assuming that, as in the case studied in Ref.~\cite{Damour:1991rq},
 it is enough to replace the second secular evolution equation in
Eqs. (\ref{secular}) by
\begin{equation} \label{edot1}
\langle\frac{d\mathbf{e}}{dt}\rangle = \mathbf{f}\times \bm{\ell} + \omega_p \, \mathbf{c}\times\mathbf{e}
\end{equation}
where $\omega_p$ describes the precession of the orbit of the Moon, which occurs with a period of $8.85$~years  \cite{book}. [Let us note in passing that the amplitude
of the eccentricity evolves according to $\frac{de}{dt}=(1-e^2)^{1/2}\mathbf{f}\cdot\mathbf{b}$.]

In addition, as the eccentricity of  the Earth-Moon system is small $e=0.0549$, we can, in first
approximation (in view of the appearance of $e$ on the r.h.s. of the evolution equation
of the angular momentum $\bm{\ell}$),
neglect the small  wobbling of the direction of the orbital angular momentum $\mathbf{c}$ and
approximate equation (\ref{edot1})  (using also $\bm{\ell} =  (1-e^2)^{1/2}\mathbf{c} \simeq \mathbf{c}$) by
\begin{equation} \label{edot2}
\langle\frac{d\mathbf{e}}{dt}\rangle = \mathbf{f}\times \mathbf{c} + \omega_p \, \mathbf{c}\times\mathbf{e}
= \omega_p \, \mathbf{c}\times (\mathbf{e} -  \mathbf{e}_{{f}})
\end{equation}
where we have introduced
\begin{equation}
\mathbf{e}_{{f}}= \frac{\mathbf{f}_\perp}{\omega_p}= \frac{3}{2}\frac{\Delta\mathbf{g}_\perp}{na \omega_p}
\end{equation}
where $\mathbf{f}_\perp = \mathbf{f} -  (\mathbf{f} \cdot \mathbf{c}) \,  \mathbf{c}$ is the component of the external force in the plane of the orbit.

The general solution of Eq.  (\ref{edot2}) reads
\begin{equation}
\mathbf{e}(t) = \mathbf{e}_{{f}}+ \mathbf{e}_p(t)
\end{equation}
where the constant vector $ \mathbf{e}_{{f}}$ describes a ``forced eccentricity'' (or a
``polarization'')  induced by the external EP-violating force, and where $\mathbf{e}_p(t)$
is the usual,  $\omega_p$-precessing free eccentricity, which is allowed in absence of external
Stark effect. Note in passing that the polarization of the elliptic orbit by the external force
is oriented in the {\it opposite} direction of the projection $\mathbf{f}_\perp \propto \Delta\mathbf{g}_\perp$  of the force on the
orbital plane. Indeed the eccentricity vector defines the direction towards the periastron,
so that the elliptic orbit is mainly elongated in the apoastron direction $-  \mathbf{e}_{{f}} \propto - \mathbf{f}_\perp$.

The final observable result, of this leading order treatment, linked to the polarization $ \mathbf{e}_{{f}}$, is a sidereal frequency
 oscillation of the Earth-Moon
range, connected with the direction   of the projection $\mathbf{f}_\perp$  of the
perturbing acceleration onto the orbital plane, of the form
\begin{equation} \label{deltarDS}
\Delta^{LO} r(t) =  \rho^{LO}_f \cos (n( t-t_0) - \phi_f)~~.
\end{equation}
where the (algebraic) amplitude is
\begin{equation} \label{rhofDS}
\rho^{LO}_f = -\frac{f_\perp a}{\omega_p}= -\frac32 \frac{\Delta g_\perp}{n \omega_p}
\end{equation}
and where  $n(t-t_0)$ is the longitude of the Moon, and
$\phi_f$ is the longitude  of the direction  $\mathbf{f}_\perp/ | \mathbf{f}_\perp|$, both longitudes
being measured within the orbital plane, from some common origin.
 [Neglecting the small inclination $i \simeq 5 $ degrees of the Moon's orbit
on the ecliptic, we can consider that both longitudes are ecliptic longitudes, counted
from the vernal equinox.]


The above result was obtained as a leading-order approximation, under the simplifying
assumption that the main effect of the solar tide on the Earth-Moon system was to
introduce a perigee precession term in Eq. (\ref{edot1}).  However, the solar tide
has more effects than this, and, as we recalled above, the theory of the lunar motion
under the combined effect of the $G(m_E + m_M)/r$ Earth-Moon potential and of the quadrupolar
tide $\frac 12 x^i x^j \partial_{ij} (G m_S/D)$ of the Sun, is a notoriously subtle theory, exhibiting
many instances of slowly converging perturbation expansions in the (not so small) expansion
parameter
\begin{equation} \label{m}
m = \frac{n'}{n-n'} \simeq 0.080849 \simeq \frac{1}{12.3687}
\end{equation}
Although the basic small parameter entering lunar theory
is the ratio of the solar tidal potential to the Earth-Moon potential, which is
 of order $m^2 \sim 10^{-2}$, the lunar perturbation theory is not an expansion
 in powers of $m^2  \sim 10^{-2}$, but proceeds (beyond the LO term) in powers of $m \sim 1/12$
 (see, e.g. \cite{BrouwerClemence}).
We note, in particular, that the perigee precession frequency of the Moon is given by
a perturbation expansion of the form
$$
\frac{\omega_p}{n} \,  = \frac{3}{2^2} \, m^2 + \frac{177}{2^5} \, m^3 + \frac{1659}{2^7} \, m^4 + \frac{85205}{2^{11}} \, m^5 + \frac{3073531}{2^{13} \cdot 3} \, m^6 + \cdots
$$
which converges so slowly that the sum of higher-order terms approximately doubles the
LO analytical result $\omega_p^{\rm LO}/n = 3 m^2/4$ (see \cite{Hill} for the literal
computation of the perturbation expansion of  $\omega_p/n$ up  to the eleventh
power of $m$).  Let us also note that the above
LO result for the range perturbation due to the Stark effect contained
$\omega_p/n = 3 m^2/4 + \cdots$ as a {\it small denominator } that significantly
amplified the effect of the external perturbing force $\mathbf{f}$. [The presence of
a small denominator is linked to the instability, recalled above, of elliptic motion
under a constant force, because this small denominator tends to zero in the limiting case of
the Lagrangian (\ref{2.1}).]

The appearance of such small denominators oblige one to tackle in a more
complete manner the Stark effect on the Moon's orbital motion. This was done
by Damour and Vokrouhlick\'{y} \cite{Damour:1995zr}
 using a  Hill-Brown treatment (with the help of a dedicated algebraic computer programme). More specifically, Ref.  \cite{Damour:1995zr} worked out the lunar range
perturbation $\Delta r(t)$ induced by an external acceleration $\mathbf{\Delta g}$ to
very high order in the powers of $m$. This range perturbation  $\Delta r(t)$ is the sum of
many different frequency components that come from the nonlinear combination of
the basic sidereal frequency $n$ of the Moon (linked to the angular distance
between the Moon and the external fixed direction $\mathbf{\Delta g}$, or rather
its projection $\mathbf{\Delta g}_{\perp}$ on the Moon's orbital plane), with even multiples
of the synodic frequency $n-n'$ linked to the angular distance (seen from the Earth)
between the Moon and the Sun. Among the spectrum of combined frequencies
$\pm (n + 2j (n-n'))$ ($j \in Z$),  two of them were found to be dominant:  the basic
sidereal frequency $\pm n$ (of period $27.32$ days), and the $j=-1$ combination $\pm (n -2 (n-n'))=\mp (n-2n')$ (of period $32.13$ days). The result of  \cite{Damour:1995zr}  can be written\footnote{
Here we change the notation of   Section IV of \cite{Damour:1995zr} : e.g. $A_G  \mathbf{N_G}
\to \mathbf{\Delta g}_\perp$,  $  S_{\rm gal}(m)  \to S_f(m)$, etc.} as
\begin{eqnarray} \label{deltarDV}
&&\Delta \, r(t) = \rho_f  \Biggl[ \cos \, [n(t-t_0)-\phi_f]  \nonumber \\
&&+ \frac{15}{8} \, m \, \frac{S'_f (m)}{S_f (m)} \cos \, [n(t-t_0) - 2 \tau(t)-\phi_f] + \ldots \Biggl]  \,.
\end{eqnarray}
Here $\tau(t) \equiv (n-n')t +\tau_0$ denotes the synodic phase, i.e. the angular distance between the Moon
and the Sun, and the overall amplitude is given by
\begin{equation} \label{rhofDV}
\rho_f=  - 2 \, \frac{\Delta g_\perp}{n'^2} \, S_f (m)
\end{equation}
where $\Delta g_\perp$ and $\phi_f$ are the magnitude, and the longitude, of the
projection  $\mathbf{\Delta g}_{\perp}$ of  the external acceleration onto the lunar orbital
plane\footnote{ Even in the full Hill-Brown treatment (with neglect of the lunar
 inclination, and of the eccentricity of the Moon's orbit), one finds that only the
 projection  $\mathbf{\Delta g}_{\perp}$ of  the external acceleration matters for
 the range perturbation.}, and
where $S_f(m)$ and $S'_f(m)$ are $m$-perturbation series that start as $1+O(m)$.
For instance, the beginning of the expansion of   $S_f(m)$ reads
\begin{eqnarray}
&S_f(m) = 1-\frac{75}{8} \, m + \frac{235}{4} \, m^2 - \frac{127 \, 637}{384} \, m^3 \nonumber \\
&+ \frac{4 \, 172 \, 299}{2304} \, m^4 + O(m^5) \,
\end{eqnarray}
and Table IV of  \cite{Damour:1995zr} gives the coefficients of this expansion to the ninth
order in $m$.  Even with such a high-order expansion one finds that the last term is
still of fractional order $10^{-3}$. This slow convergence is related to the presence of
a pole in the series  $S_f(m)$ and $S'_f(m)$ near $m_{\rm cr} \simeq - 0.184 07$.  To get
an accurate numerical estimate of  these series  Ref.~ \cite{Damour:1995zr}
used a Pad\' e resummation, with the results [for $m=m_{\rm Moon}$ given by Eq. (\ref{m})]
$$
S_f (m) \simeq 0.5050
$$
and
$$
\frac{15}{8} \, m \, \frac{S'_f (m)}{S_f (m)} \simeq \frac{1}{5.94}
$$
for the fractional coefficient of the subleading term with frequency $n-2(n-n')= -(n-2n')$.

An observationally important aspect of the result  (\ref{deltarDV}) is the appearance of a
specific combination of two  harmonics, with known periods and phase, and with nearly
comparable magnitudes. In particular, the fact that the amplitude of the  $n-2n'$
harmonic is only $\simeq 6$ times smaller than the LO $n$ harmonic is a result
of the subtleties of lunar perturbation theory.  This term comes from the basic
solar tide perturbation which is proportional to $m^2$, but it has been amplified to
the $O(m)$ level  by a small denominator
(with the additional factor $15/8 \simeq 2$, leading to $15 m/8 \simeq 1/6$).
We have checked the presence of this subleading term
 by directly solving the {\it forced}  Hill's equation  \cite{Hill,BrouwerClemence,Damour:1995gi} $d^2 q(\tau)/d\tau^2 + \Theta(\tau) q(\tau)=\sigma(\tau)$ for
 the transverse perturbation $ q(\tau)$ to Hill's variational orbit. Here,
 $\tau= (n-n') t +\tau_0$ as above. In that formulation,  the frequency component $q_j \exp[ i(1+m+2j) \tau]$
 comes with the  denominator $\theta_0 - (1+m +2j)^2 $ which is small  when $ j=-1$,
 being $O(m)$  [namely, $ \theta_0 - (m-1)^2=  [1+ 2m +O(m^2)] -(m-1)^2 = 4m + O(m^2)$].

 For what concerns the leading term, with frequency $n$,  in Eq. (\ref{deltarDV}), it corresponds
 to the result   Eq. (\ref{deltarDS}) of the approximate treatment explained above. In both
 cases $\phi_f$ is the longitude of the external acceleration projected within the
 orbital plane. Note that if one were using the LO analytical results
 $(\omega_p/n)^{LO} = (3/4) m^2$ and $S_f^{LO}(m)=1$, Eq. (\ref{rhofDS}) would read
 $-2 \Delta g_\perp/n'^2$ and would agree with Eq. (\ref{rhofDV}).
 However, the exact result  Eq. (\ref{rhofDV})  is smaller
 than this by about  a factor two, because of the correcting factor $S_f (m) \simeq 0.5050$.
 As noted in Ref.~ \cite{Damour:1995zr},
 when using in Eq.  (\ref{rhofDS}) the actual
 perigee precession $\omega_p$ (which is about twice larger than its LO estimate
 $(\omega_p/n)^{LO} = (3/4) m^2$), one captures most of the effect of the slowly
 converging series  $S_f(m)$.

We can finally apply the result  Eq. (\ref{deltarDV}) to the EP-violating acceleration
$\Delta \mathbf{g} = \mathbf{g}_M - \mathbf{g}_E$ where each $\mathbf{g}_A$ is given by equation (\ref{gA}). Let us first note that the result only depends on the amplitude
and longitude of the projection on the orbital plane of the  vectorial sum of  the external accelerations
\begin{equation} \label{gME}
\Delta \mathbf{g} = \mathbf{g}_M - \mathbf{g}_E=
 - \sum_i (Q_{r_i}(M)- Q_{r_i}(E))~B_{r_i} c^2~\mathbf{\hat{z}}_{r_i}.
\end{equation}
Alternatively, one could write the range perturbation as a sum of terms
of the type of the r.h.s. of Eq. (\ref{deltarDV}), each one having an
amplitude $\rho_f(r_i)$ and a phase $\phi_f(r_i)$. [Note that we consider here
algebraic amplitudes (that can be negative), and that the longitudes $\phi_f(r_i)$
always refer to the direction of the projection $\mathbf{\hat{z}}_{r_i \, \perp}$
of the gradient direction $ + \mathbf{\hat{z}}_{r_i}$.]
Let us, as it simplifies the writing of
our results, make the natural assumption that all the cosmological gradients
of the coupling constants are parallel to each other, i.e.
$  \mathbf{\hat{z}}_{r_i} = \mathbf{\hat{z}}$ independently of the label $r_i$,
and let us denote by $\phi_f$ the longitude of the projected gradient direction
$\mathbf{\hat{z}}_\perp$. This leads to a total range perturbation of the form
Eq. (\ref{deltarDV}), with a total (algebraic) amplitude of the form (after
cancellation of two minus signs)
\begin{equation}
\rho_f^{tot} = 2 S_f (m)   \sum_i (Q_{r_i}(M)- Q_{r_i}(E))~\frac{ B_{r_i \perp} c^2}{n'^2}
\end{equation}
where $B_{r_i \perp}$ is the magnitude of the projected gradient
$B_{r_i } \hat{z}_\perp$.

Note that the numerical prefactor $2 S_f (m) \simeq 1.010$ is close to $1$, and that the
parameter combination $ B_{r_i \perp} c^2/n'^2$, where we recall that $n'= 2 \pi/ 1 {\rm yr}$,
is the product of an acceleration by the square of a time, and is indeed a length. [Alternatively,
we can think of it as the product of the spatial gradient  $ B_{r_i \perp} = [{\rm length}]^{-1}$,
by the squared length $c^2/n'^2=(1 \, {\rm lyr})^2/ 4 \pi^2$].
As a first orientation, note that a gradient of order $B\sim 10^{-6} \, {\rm Glyr}^{-1}$, i.e.
$ B c^2 \sim 10^{-12} \, {\rm cm/s}^2$ (similar to
 the recent suggestions, Eqs. (\ref{Balpha}), (\ref{Bmu}), (\ref{Balphac2}), (\ref{Bmuc2})) corresponds to
a figure of  merit  $ B c^2/n'^2 \simeq 25$ cm. In order to estimate the corresponding
signal in lunar laser ranging, we need next to estimate the numerical value of the various
EP-violating charge differences $Q_{r_i}(M)- Q_{r_i}(E)$ corresponding to the difference
in composition of the Moon and the Earth. This will be the focus of the next Section.

Before tackling this issue, let us briefly mention that the central values of
the equatorial coordinates $\alpha=261$ degrees, $\delta=- 58$ degrees of
the cosmological gradient of the fine structure constant reported in \cite{Webb:2010hc}
corresponds to an ecliptic latitude equal to $\beta=- 34.7$ degrees, and an
ecliptic longitude equal to $\lambda= - 95.8$ degrees. The latter ecliptic longitude
predicts\footnote{As above, we are here neglecting the small inclination of
the Moon's orbit on the ecliptic. Note that taking into account this inclination,
and its secular variation, would predict an additional small adiabatic  variation
of the amplitude $\rho_f$ (and phase $\phi_f$) of the range perturbation $\Delta r(t)$,
corresponding to the secular wobbling of the lunar orbital plane.}
the value of the longitude entering the range perturbation Eq. (\ref{deltarDV}),
namely $\phi_f=\lambda$. On the other hand, the ecliptic latitude $\beta$ enters
the observable range $\rho_f$ through the projection of the cosmological gradient direction
onto the orbital plane, i.e.(essentially) onto the ecliptic. More precisely we have
$B_{r_i \perp}= B_{r_i} \cos \beta$. Note that $\cos \beta = 0.822$ for the gradient
reported by Webb et al. so that this projection (that we shall include in the estimates of the
next Section) reduces only by $18 \%$  the full possible
observable effect of such a gradient on the lunar motion.


\section{EP violating charges}

 Let us now estimate the numerical values of the various EP-violating charge differences
 $Q_{r_i}(M)- Q_{r_i}(E)$ entering the magnitude of the cosmologically induced Earth-Moon
 differential acceleration (\ref{gA}).  This issue has been discussed in our previous work
 \cite{Damour:2010rm,Damour:2010rp}  where we described the leading dependence of atomic masses on the parameters of the Standard Model. Because of the large neutron and proton masses, the dominant determining factor for
all atomic masses is simply the scale of the strong interactions $\Lambda_{QCD}$. However,
as already mentioned above, the main dependence of atomic masses on $\Lambda_{QCD}$  is universal, and is
conveniently factored out by rewriting each atomic mass $M_A$ as
$M_A= \Lambda_{QCD} f(r_i)$, where the $r_i$'s (for $i \ne 0$)
denote the dimensionless ratios Eq. (\ref{defri}). As a consequence, the charge $Q_{r_0}$
associated to $r_0= \Lambda_{QCD}/M_P$ via the general definition  Eq. (\ref{defQ})
is simply  $Q_{r_0}(A) \equiv 1$, and the only non-zero contributions to the differential
acceleration (\ref{gA}) will come from the
 dependence of $M_A/ \Lambda_{QCD} $  on the masses of the light quarks, on the electron mass and on the electromagnetic fine structure constant. For the quark masses, instead of considering separately the masses of the up and down quarks, it is convenient to consider their
 average and difference, namely
\begin{equation}
\hat{m} = (m_d+m_u)/2,~~~~\delta m = (m_d-m_u)
\end{equation}
and to work with the fine structure constant $\alpha$ together with the dimensionless ratios
\begin{equation} \label{massratios}
r_{\hat{m}} = \frac{\hat{m}}{\Lambda_{QCD}},~~ r_{\delta {m}} = \frac{\delta{m}}{\Lambda_{QCD}},~~ r_{m_e} =\frac{m_e}{\Lambda_{QCD}} ~~.
\end{equation}
Because the fermion masses are the product of a dimensionless Yukawa coupling $\Gamma_i$ and the Higgs vacuum expectation value (vev) $v$, $m_i = \Gamma_i v/\sqrt{2}$, these ratios can be considered as the product of the dimensionless $\Gamma_i$ by the ratio $v/ \Lambda_{QCD} $
between the basic weak-interaction scale $v$ and the basic strong-interaction scale
$\Lambda_{QCD} $.  In view of the independence of the mechanisms leading to the
appearance of the two basic scales $v$ and $\Lambda_{QCD} $, it seems a priori theoretically natural to expect that the cosmological gradients (if any) of the three mass ratios
(\ref{massratios}) will be of similar magnitudes, and therefore similar to that of the
ratio $\mu = m_e/m_p$ for which the value (\ref{Bmu}) has been recently suggested.


Refs.~ \cite{Damour:2010rm,Damour:2010rp} derived the folllowing approximate estimates
for the four effective charges   $Q_{r_i}$ associated to the three mass ratios (\ref{massratios}),
and to $\alpha$:
\begin{eqnarray}
\label{Qmhat}
 Q_{\hat m} &=& F_A \left[ 0.093 -\frac{0.036}{A^{1/3}}- 0.020 \frac{(A-2Z)^2}{A^2} \right. \nonumber \\
  &~&~~~~\left.~~
- 1.4 \times 10^{-4} \, \frac{Z(Z-1)}{A^{4/3}} \right]  ,
\end{eqnarray}
\begin{equation}
\label{Qdeltam}
Q_{\delta m} =  F_A \left[0.0017   \, \frac{A-2Z}{A} \right] ,
\end{equation}
\begin{equation}
\label{Qme}
 Q_{m_e} = F_A \left[ 5.5 \times 10^{-4}  \, \frac{Z}{A} \right] ,
\end{equation}
and
\begin{equation}
\label{Qe}
Q_\alpha = F_A \left[  -1.4 + 8.2 \frac{Z}{A} + 7.7 \frac{Z(Z-1)}{A^{4/3}}  \right]\times 10^{-4}.
\end{equation}
where $F_A \equiv A m_{\rm amu}/M_A$, with $m_{\rm amu} = 931$ MeV denoting the
atomic mass unit. The common factor $F_A$ is very close to 1, and we shall replace it
by 1 in our estimates below.

Approximating the Moon  as made of silicate (i.e. essentially SiO$_2$),
and the Earth  as made of a mantle of silicate and a core of iron (representing 32\% of its
mass), the above formulas yield the following Moon-Earth charge differences

\begin{eqnarray}
Q_{\hat m}(M) -  Q_{\hat m}(E) &=&  - 1.1 \times 10^{-3} \\
Q_{\delta m}(M) -  Q_{\delta m}(E) &=& - 3.8 \times 10^{-5} \\
Q_{ m_e}(M) -  Q_{m_e}(E) &=&  + 6.1 \times 10^{-6}     \\
Q_{\alpha}(M) -  Q_{\alpha}(E) &=&  - 3.1 \times 10^{-4}
\end{eqnarray}

We see that the most important
charges for EP violation are $Q_{\hat{m}}$ and $Q_\alpha$.
This dominance of   $Q_{\hat{m}}$ and $Q_\alpha$ over the other charges is true
 for most values of $Z,~ A$. It is related both to the rather small coefficients entering
 $Q_{\delta m}$, Eq. (\ref{Qdeltam}), and $ Q_{m_e} $, Eq. (\ref{Qme}), and to the
 fact that $A \simeq 2 Z$ along the periodic table. If we use the facts that
 $F_A -1 =O(10^{-3})$, and that $A \simeq 2 Z$,
 we can simplify the composition dependence of the main EP-violating charges
 $Q_{\hat{m}}$ and $Q_\alpha$ as $Q_{\hat{m}} \simeq Q'_{\hat m} +$cst. and
 $Q_\alpha \simeq Q'_\alpha +$cst., where \cite{Damour:2010rm,Damour:2010rp}
\begin{eqnarray}
Q'_{\hat m} &= -\frac{0.036}{A^{1/3}} - 1.4 \times 10^{-4} \, \frac{Z(Z-1)}{A^{4/3}}  \\
 Q'_\alpha &= + 7.7 \frac{Z(Z-1)}{A^{4/3}}  \times 10^{-4}
 \end{eqnarray}
We list the values of $-Q'_{\hat m}$ and $ Q'_\alpha$ for a sample of materials in Table I.
This list shows that the maximum value of a charge difference would be the
$Q_{\hat{m}}$ difference between a heavy element and a light one with
$Q_{\hat{m}}({\rm heavy}) - Q_{\hat{m}}({\rm light}) \simeq + \, 10^{-2}$.
The corresponding  maximum acceleration level
$\Delta g_{\hat m}^{\rm max} = \Delta Q_{\hat{m}}^{\rm max} B_{\hat m} c^2 \simeq 10^{-2} B_{\hat m} c^2$
would numerically be
\begin{equation}
\Delta g_{\hat m}^{\rm max} \simeq  \left(\frac{B_{\hat m}}{10^{-6} ~ {\rm Glyr}^{-1}}\right)  \times 10^{-14} {\rm cm/s}^2  \nonumber
\end{equation}
for a cosmic  gradient of order of those reported by Webb et al.
\begin{table}[h]\centering
\caption{Approximate EP-violating `effective charges' for a sample of materials. These charges are averaged over the (isotopic or chemical, for SiO$_2$) composition.}
\begin{tabular}{ccccc}
\\
{\rm Material} &$A$ &$Z$ &$-Q'_{\hat m}$ &$Q'_\alpha$ \\ \\
{\rm Li} &7 &3 &18.88 $\times 10^{-3}$ &0.345 $\times 10^{-3}$ \\
{\rm Be} &9 &4 &17.40 $\times 10^{-3}$ &0.494 $\times 10^{-3}$ \\
{\rm Al} &27 &13 &12.27 $\times 10^{-3}$ &1.48 $\times 10^{-3}$ \\
{\rm Si} &28.1 &14 &12.1 $\times 10^{-3}$ &1.64 $\times 10^{-3}$ \\
{\rm SiO$_2$} &... &... &13.39 $\times 10^{-3}$ &1.34 $\times 10^{-3}$ \\
{\rm Ti} &47.9 &22 &10.28 $\times 10^{-3}$ &2.04 $\times 10^{-3}$ \\
{\rm Fe} &56 &26 &9.83 $\times 10^{-3}$ &2.34 $\times 10^{-3}$ \\
{\rm Cu} &63.6 &29 &9.47 $\times 10^{-3}$ &2.46 $\times 10^{-3}$ \\
{\rm Cs} &133 &55 &7.67 $\times 10^{-3}$ &3.37 $\times 10^{-3}$ \\
{\rm Pt} &195.1 &78 &6.95 $\times 10^{-3}$ &4.09 $\times 10^{-3}$ \\
\end{tabular}
\end{table}

\section{Observational signals linked to possible cosmological gradients}

Putting together our results, keeping only the dominant terms linked to  $Q_{\hat{m}}$ and $Q_\alpha$, and scaling the possible cosmological gradients of
${\hat m}$ and $\alpha$ by the recently reported values\footnote{As mentioned above,
it is natural to expect that a cosmological gradient of $\mu=m_e/m_p$ implies a similar gradient
in the weak-interaction/strong-interaction ratios $r_i$, and notably in ${\hat m}/\Lambda_{QCD}$.} Eqs. (\ref{Bmu}), (\ref{Balpha})
[and  (\ref{Bmuc2}), (\ref{Balphac2})], we conclude that those cosmological gradients
entail EP-violating differential accelerations on the Moon directed along the unit
vector $\mathbf{\hat{z}}_\perp$ (with ecliptic longitude equal to $\phi_f \simeq \lambda \simeq - 95.8$ degrees), and with algebraic magnitudes
\begin{equation}
\Delta g_\perp = \Delta g_{\hat m \perp} + \Delta g_{\alpha \perp}
\end{equation}
where
\begin{eqnarray} \label{ghatm}
 \Delta g_{\hat m \perp} &= & -  (Q_{\hat m}(M)- Q_{\hat m}(E))~B_{\hat m \perp} c^2  \\
 &= &
 + 2.2  \left(\frac{B_{\hat m}}{2.6 \times 10^{-6} ~{\rm Glyr}^{-1}}\right)  \times 10^{-15} {\rm cm/s}^2  \nonumber
 \end{eqnarray}
 and
\begin{eqnarray} \label{galpha}
 \Delta g_{\alpha \perp} &=  &  -  (Q_{\alpha}(M)- Q_{\alpha}(E))~B_{\alpha \perp} c^2 \\
 &= &
 + 2.7  \left(\frac{B_{\alpha}}{1.1 \times 10^{-6} ~{\rm Glyr}^{-1}}\right) \times 10^{-16} {\rm cm/s}^2 \nonumber
 \end{eqnarray}
These differential accelerations then induce a perturbation in the Earth-Moon range
which has the specific time signature (\ref{deltarDV}) with an algebraic magnitude
\begin{equation}
\rho_f = \rho_{\hat m} + \rho_{\alpha}
\end{equation}
where
\begin{equation} \label{rhom}
\rho_{\hat m} = - \left(\frac{B_{\hat m}}{2.6 \times 10^{-6} ~{\rm Glyr}^{-1}}\right) 0.59 \, {\rm mm}
\end{equation}
and
\begin{equation} \label{rhoalpha}
\rho_{\alpha} = - \left(\frac{B_{\alpha}}{1.1 \times 10^{-6} ~{\rm Glyr}^{-1}}\right) 0.068 \, {\rm mm}
\end{equation}

As we see, the dominant effect is expected to be linked to the cosmological gradient
of $\hat m/\Lambda_{QCD} $, so that the recent findings of
Webb and collaborators \cite{Webb:2010hc,Berengut:2010yu} suggest the presence
of millimeter-level sidereal fluctuations in the Earth-Moon range. Such fluctuations, if
they exist, should be detectable by the APOLLO experiment. Indeed, this experiment
has shown its capability of obtaining ``normal-point'' range
measurements with nightly median uncertainty of 1.8 mm for their entire data set,
and 1.1 mm for their recent data \cite{apollo09}. By accumulating
millimeter-level range data over a sufficiently long time (comparable to the perigee
period $\simeq 8.85$ yr), the APOLLO experiment should be able both to
decorrelate the specific sidereal signal  (\ref{deltarDV}) from the many existing
Newtonian range effects (which include  synodic, $ n-n'$, and anomalistic, $\omega_r=n-\omega_p$, frequencies),
and to measure its amplitude $\rho_f$ to a fraction of a millimeter. Depending on
the result of such an analysis, the APOLLO experiment could either establish the
reality of a cosmological gradient of coupling constants, or set upper bounds on the
gradients $B_{\hat m}$ and $B_\alpha$ (or more precisely on the combination
 $B_{\hat m \perp} + 0.28 B_{\alpha \perp}$ entering $\Delta g_\perp$) at levels smaller than the
 levels Eqs. (\ref{Bmu}), (\ref{Balpha}) [and  (\ref{Bmuc2}), (\ref{Balphac2})].

 Let us mention in this respect that   a sidereal range perturbation
 of the approximate form (\ref{deltarDS}), with a phase $\phi_f$ linked
 to the center of the Galaxy, has been searched for in the pre-APOLLO,
 few-centimeter-level LLR data after the suggestion of Ref. \cite{Nordtvedt94}.
 Nordtvedt, M\"uller and Soffel \cite{NordvedtMullerSoffel} published an upper limit of
 \begin{equation}
\Delta g_{\rm gal} < 3 \times 10^{-14} {\rm cm/s}^2
\end{equation}
on a possible perturbing differential acceleration linked to the Galactic center,
while a further analysis of M\"uller (cited in \cite{Muller:2005vv}) obtained
\begin{equation}
\Delta g_{\rm gal} = (4 \pm 4) \times 10^{-14} {\rm cm/s}^2
\end{equation}
The gain in sensitivity of the APOLLO experiment, by more than a factor 10, and the richer
time signature of the more complete signal (\ref{deltarDV}), make us expect
that it will  be possible to probe the acceleration levels (\ref{ghatm}),  (\ref{galpha}),
above.

It is instructive to compare the (potential)
sensitivity of LLR experiments to external EP-violating
accelerations with the sensitivity of other EP experiments.
There have  been constraints on anomalous cosmic accelerations  by laboratory based EP tests. The most precise
is quoted as a differential acceleration in any direction of the sky \cite{Schlamminger:2007ht}
$$
 | g({\rm Be}) - g({\rm Ti}) | < 8.8 \times 10^{-13} ~{\rm cm/s}^2 \,  (95 \% \, C.L.)
$$
The charge differences $Q_{r_i}({\rm Be}) - Q_{r_i}({\rm Ti}) $ are again dominated
by   $Q_{\hat m}({\rm Be}) - Q_{\hat m}({\rm Ti}) = - 7.23 \times 10^{-3}$
and $Q_{\alpha}({\rm Be}) - Q_{\alpha}({\rm Ti}) = - 1.56 \times 10^{-3}$.
Assuming, as above, that the effect of the gradient of $\hat m$ (which couples
to the dominant charge difference) dominates, the above upper bound
on $ | g({\rm Be}) - g({\rm Ti}) | $
can be readily converted to a bound on the corresponding cosmological gradient $B_{\hat m}$,
namely
\begin{equation} \label{labbound}
| B_{\hat m} |  <   1.3 \times 10^{-4} ~{\rm Glyr}^{-1}
\end{equation}
This is is weaker than the recently suggested (theoretically similar) gradient Eq. (\ref{Bmu}) by
a factor $\simeq 50$. Such a difference in {\it  acceleration sensitivity} between Earth-based
EP experiments and LLR ones might seem surprising in view of the fact that both types of experiments currently lead to comparable limits on the (E\"otv\"os) EP-violation
parameter $\eta = \Delta g/g$, namely
$ \eta_{\rm Earth Moon} = (-1.0 \pm 1.4) \times 10^{-13}$ \cite{lunarlaser},
versus $ \eta_{\rm Be Ti} = (0.3 \pm 1.8) \times 10^{-13}$  \cite{Schlamminger:2007ht},
and that both types of experiments use comparable background accelerations $g$ in
the ratio $\Delta g/g$. [Indeed, the $g$ due to the Sun at Earth is $g_S \simeq 0.6 $ cm/s$^2$,
while torsion balance experiments use only the horizontal component of the Earth gravity,
namely
$g_{E \, \perp} \simeq 1.7 $ cm/s$^2$ at a latitude of $45$ degrees.]
We note that  the greater sensitivity of LLR experiments to external (especially
fixed-direction) accelerations is essentially rooted in the specific
Stark instability mentioned above.
Indeed, generally speaking,
a differential acceleration $\Delta g$ acting during a characteristic time $t_c$ (which
is $t_c \sim \omega^{-1} = T/(2 \pi)$ for a periodic phenomenon of angular
frequency $\omega$ and period $T$) corresponds to a measurable displacement
of order $\Delta r \sim \Delta g t_c^2 = \Delta g/\omega^2$. In the LLR case, we
saw above that the  range perturbation is $\Delta r \sim \Delta g/n'^2$
which is {\it larger} than the expected perturbation $\sim \Delta g/n^2$ associated
to the lunar frequency $n$ by a factor
$$
\left(\frac{n}{n'}\right)^2 = \left(\frac{1 {\rm year}}{1 {\rm sidereal \, month}}\right)^2=(13.37)^2=178.7
$$
This amplification factor lies at the root of the increased sensitivity of LLR experiments
to external EP-violating  accelerations having a fixed direction. We note in passing
that the LLR sensitivity to the usually considered Laplace-Nordtvedt solar-rooted
EP-violating acceleration is only amplified, w.r.t. $ \Delta g/n^2$, by a parametrically
smaller factor $\sim (3/2) (n/n') \sim 20$, i.e. about ten times less than in the ``Stark'',
fixed direction case. This difference is due to the difference in the corresponding
nearly resonant denominators, namely, in Hill's equation, a denominator
 $ \theta_0 - (1+m)^2=  [1+ 2m +O(m^2)] -(1+m)^2=O(m^2)$ in the sidereal-frequency (Stark) case,
 versus  $ \theta_0 - (1)^2=  [1+ 2m +O(m^2)] -1=2m + O(m^2)$ in the synodic-frequency (Laplace-Nordtvedt) one.

 Let us finally note that presently planned improved EP tests such as the Satellite Test
 of the Equivalence Principle (STEP) \cite{step} ($\eta \sim 10^{-18}$) or  proposed
 cold-atom-technology tests  (($\eta \sim 10^{-17}$)  \cite{Dimopoulos:2006nk},
 which will make use of the full strength of the Earth gravity,  $g_E \simeq 980 \, {\rm cm/s}^2$,
 will be able to probe the cosmological-gradient-induced
 differential accelerations discussed above. Indeed, the acceleration (\ref{ghatm})
 linked to a cosmic gradient of $\hat m$ can be rewritten (modulo a cosine factor,
 with a specific time dependence
 linked to the projection onto the sensitive direction of the considered EP experiment) as
 $$
\eta =\frac{\Delta g}{g} = 2.5 \,  \frac{g_E}{g} \left( \frac{\Delta Q_{\hat m}}{10^{-2}} \right) \left(\frac{B_{\hat m}}{2.6 \times 10^{-6} ~{\rm Glyr}^{-1}}\right)  \times 10^{-17}
$$
where we allowed the considered man-made EP test to optimize the choice of materials
by having a $\Delta Q_{\hat m} \sim 10^{-2}$, i.e. ten times better than for the Earth-Moon case
(see Table I).
This result shows that
the LLR test of the cosmological acceleration  (\ref{ghatm}), that should be doable
by the APOLLO experiment,  corresponds (from the point of view
of the sensitivity to a cosmic gradient) to an Earth-based
EP test at the $\eta \sim 10^{-17} $ level.

\section*{Acknowledgements}
We thank Eric Adelberger for useful information about the APOLLO experiment.
JFD thanks the IHES for hospitality during this project and also acknowledges partial support by the NSF grants PHY - 0855119 and by the Foundational Questions Institute.

\end{document}